\documentclass[a4paper,11pt]{article}
\pdfoutput=1 % if your are submitting a pdflatex (i.e. if you have
\usepackage{jcappub} % for details on the use of the package, please
\usepackage[T1]{fontenc} % if needed

\title{Mutual information between galaxy properties and the initial predisposition}

\author[a,b]{Jun-Sung Moon}
\author[a,1]{, Jounghun Lee \note{Corresponding author.}}

\affiliation[a]{Department of Physics and Astronomy, Seoul National University, \\
Kwanak-ro 1, Kwanak-gu, Seoul 08826, Republic of Korea}
\affiliation[b]{Research Institute of Basic Sciences, Seoul National University,  \\
Kwanak-ro 1, Kwanak-gu, Seoul 08826, Republic of Korea}

\emailAdd{jsmoon.astro@gmail.com}
\emailAdd{cosmos.hun@gmail.com}

\abstract{The immense diversity of the galaxy population in the universe is believed to stem from their disparate merging and star formation histories, and multi-scale 
influences of diverse environments.  No single causal factor of the initial state is known to explain how the galaxies formed and evolved to end up 
possessing such various traits as they have at the present epoch. However, several observational studies have revealed that the key physical properties of the observed 
galaxies in the local universe appeared to have a much simpler, lower-dimensional correlation structure than expected, the origin of which remains unexplained. 
Speculating that the emergence of such a simple correlation structure of the galaxy properties must be triggered by nature rather than by nurture, 
we explore if the present galaxy properties may be correlated with the initial precondition for protogalaxy angular momentum, $\tau$, and test it against the data 
from the IllustrisTNG300-1 hydrodynamic simulation. Employing Shannon's information theory, we discover that $\tau$ shares a significantly large amount of mutual information 
with each of the four basic traits of the TNG galaxies at $z=0$: the spin parameters, formation epochs, stellar-to-total mass ratios, and fraction of kinetic energy in ordered rotation. 
These basic traits except for the stellar-to-total mass ratios are found to contain even a larger amount of MI about $\tau$ than about the total masses and environments 
for the case of giant galaxies with $11.5\le \log[M_{\rm t}/(h^{-1}\,M_{\odot})]<13$.  Our results imply that the initial condition of the universe must be more impactful on the galaxy 
evolution than conventionally thought.}
\begin{document}
\maketitle
\flushbottom

\section{Introduction}\label{sec:intro}

The basic building blocks of the universe are the galaxies whose formation and evolution mechanisms have yet to be fully comprehended~\cite{GT_review}.  
The scaling relations among their basic traits are one of those observable features that can provide us with crucial clues to the major key players in the act of galaxy evolution, 
identification of which in turn can allow us to constrain theoretical models for the galaxy formation ~\cite[][and references therein]{rob-etal06,cou-etal07}. 
In the hierarchical structure formation paradigm based on the concordance cosmology where the cosmological constant $\Lambda$ and cold dark matter 
(CDM) are the most dominant energy and matter contents, respectively~\cite{mill}, the existence and structure of mutual correlations among various galaxy properties 
are believed to be established by multiple factors including the total mass, virial conditions~\cite{GG72,PS74,dav-etal85,bbks86}, multi-scale influences of cosmic web 
environments~\cite{web96} on which the hierarchical merging rates, star formation and quenching, and feedbacks are dependent~\cite[e.g.,][]{got-etal01,kau-etal04,hir-etal14}. 
In other words, the standard paradigm requires multiple independent parameters to describe the scaling relations which reflect the highly nonlinear evolutionary tracks of the 
galaxies~\cite[][and references therein]{nai-etal10}. 

Several observational studies, however, revealed that the full correlations among the basic galaxy properties are in fact redundant.  Disney {\it et al.}~\cite{dis-etal08} conducted a principal 
component analysis (PCA) of the five basic traits of $195$ HI-selected galaxies from the blind $21$cm surveys, and demonstrated that all of the five basic traits exhibited strong linear 
correlations exclusively with the first PC, being almost independent of the other four PC's ~\cite[see also][]{gar-etal09}. Given that the total masses of the HI galaxies used in the analysis 
of~\cite{dis-etal08} lie in a narrow range, it was difficult to attribute their results to the mass dependence of galaxy properties.
Although the identity of the first PC was not revealed by their PCA analysis, it was suspected by ~\cite{dis-etal08} that the existence of such a low-dimensional correlation structure 
of the galaxy properties would challenge hierarchical structure formation scenario based on the CDM paradigm, which predict that the evolution histories should differ 
among individual galaxies, depending on multiple factors. 
Later,~\cite{cha-etal12} applied the same PCA-based method to a larger sample of the HI galaxies and confirmed the results of~\cite{dis-etal08}, although they were more cautious about 
interpreting their results as a counter-evidence against the CDM. 
Very recently,~\cite{sha-etal23} performed a meticulous PCA analysis of the galaxy optical spectra from the Sloan Digital Sky Survey, separately treating the quiescent, 
star-forming, and active galactic nuclei galaxies.  Their results turned out to be in line with the dominance of the first PC, which~\cite{dis-etal08} already detected but failed 
to identify.

Recall that among several different galaxy properties, what exhibited the strongest linear correlation with the first PC in the analysis of~\cite{dis-etal08} was the galaxy sizes, 
an observational proxy of the galaxy spin parameters~\cite{jim-etal98,KL13,kul-etal20,per-etal22}.  The production mechanism of such a low-dimensional correlation 
structure may well lurk in the initial conditions of the universe rather than in the evolutionary paths of the galaxies.  
In fact, several numerical works already revealed that the basic properties of present galaxies seemed to be closely linked with the origin of their protogalactic 
angular momenta~\cite[][and references therein]{cod-etal15,cad-etal22,ML24}.  

The protogalactic sites that are believed to coincide with the initial density peaks~\cite{GG72,PS74,dav-etal85,bbks86} are exposed to the tidal torque effects of the surrounding 
anisotropic matter distribution until the turn-around moments~\cite{dor70,whi84}. These protogalactic tidal interactions lead to the generation of protogalaxy angular momentum 
at first order under the precondition that their inertia tensors are misaligned with the initial tidal tensors~\cite{whi84,lp00,lp01,por-etal02}.  
In our companion paper~\cite{ML24}, we have shown that the strength and tendency of galaxy spin alignment depend strongly on the degree of the misalignment between the 
protogalactic inertia and initial tidal fields (say, $\tau$).  In this paper, we will investigate if and how strongly the basic properties of the present galaxies depend on $\tau$. 

The outline of the upcoming sections of this paper is as follows: 
In section~\ref{sec:tau}, the numerical data is described and the initial precondition for protogalaxy angular momentum is quantified. 
In section~\ref{sec:trend}, the numerical results are presented on the trends by which the ensemble averages of the key galaxy traits over the controlled samples vary with 
the initial precondition. 
In section~\ref{sec:MI}, the results are presented on the amounts of mutual information between the initial precondition and the basic traits of the galaxies.
In section~\ref{sec:con}, the main results are summarized, their implications are discussed, and a bottom line is drawn up.

\section{Impact of the initial precondition on the basic galaxy traits}

\subsection{Measurements of the initial precondition, $\tau$}\label{sec:tau}

We exploit the halo catalogs as well as the particle snapshots from the 300 Mpc volume run (TNG300-1) of the IllustrisTNG suite of cosmological gravo-magneto-hydrodynamical simulations~\citep{tng1,tng2,tng3,tng4,tng5, tng6} for the Planck $\Lambda$CDM cosmology~\cite{planck15}.  
The TNG300-1 simulation run accommodated a total of $2500^{3}$ baryonic cells and an equal number of DM particles as massive as $1.1$ and $5.9$ in unit of 
$10^{7}\,M_{\odot}$, respectively, in a periodic cubic box of a side length $205\,h^{-1}{\rm Mpc}$. 
From the TNG300-1 catalog of the friends-of-friends (FoF) groups at $z=0$, we extract the substructures identified by the SUBFIND algorithm~\cite{subfind} to 
prepare three main samples. 

The most massive substructure of each FoF group is referred to as a central, while the less massive ones are called the satellites. 
Only the former is included in the main samples, since the latter is liable to complicated nonlinear processes like dynamical friction, tidal heating 
and mass loss inside their host groups~\cite[e.g.,][]{TB01}, which are likely to undermine the $\tau$-dependence of their properties.  
Placing the lower limit number cut, $300$, on the constituent particles of the central substructures~\cite{bet-etal07}, we create the main samples of 
the central substructures whose total masses, $M_{\rm t}$, fall in three different logarithmic ranges: $11.5\le m_{t} \le 13$,\  $11\le m_{t}\le 11.5$ 
and $10.5\le m_{t}\le 11$ with $m_{t}\equiv \log\left[M_{\rm t}/(h^{-1}\,M_{\odot})\right]$.  The central substructures belonging to the three main samples, 
whose numbers are listed in the second column of table~\ref{tab:ng}, will be referred to as the galaxies hereafter. 

Applying the same algorithm as described in our prior work~\citep{ML23,ML24} to the spatial distributions of DM particles at the initial redshift of the TNG300-1 
simulation, $z_{\rm ini}\equiv 127$, we construct the initial tidal field, ${\bf T}_{\rm ini}$, on $512^{3}$ grid points, filtered with a Gaussian window function on the 
comoving scale of $R_{f}$ by taking the following steps: 
(i) Constructing the initial density field, $\delta_{\rm ini}$, at $z_{\rm ini}$ by applying the cloud-in-cell method to the particle snapshot at $z_{\rm ini}$. 
(ii) Performing a fast Fourier transformation (FFT) of $\delta_{\rm ini}$ and computing the Gaussian smoothed initial tidal field in Fourier space, 
$\tilde{\bf T}_{\rm ini}\equiv (\tilde{T}_{{\rm ini},\alpha\beta})$, 
as  $\tilde{T}_{{\rm ini},\alpha\beta}({\bf k})\equiv k_{\alpha}k_{\beta}\tilde{\delta}_{\rm ini}({\bf k})\exp(-\vert{\bf k}\vert^{2}R^{2}_{f}/2)/\vert{\bf k}\vert^{2}$ 
with wave vector ${\bf k}=(k_{\alpha})$.
(iii) Performing an inverse FFT of $\tilde{\bf T}_{\rm ini}$ to obtain the real-space initial tidal field ${\bf T}_{\rm ini}=({T}_{{\rm ini},\alpha\beta})$. 
 
We set $R_{f}$ at $8\bar{r}_{\rm vir}$ where $\bar{r}_{\rm vir}$ is the mean virial radius of the galaxies in each main sample, recalling the fact that 
the influence of the initial tidal field on a protogalaxy with total mass $M_{t}$ becomes maximum on the scale equivalent to twice the protogalaxy Lagrangian radius, 
which in turn is approximately $8$ times the galaxy virial radius~\cite{ML24}. 
The evolved tidal field at $z=0$, ${\bf T}_{0}(R_{f})$, is also created on the $512^{3}$ grids by following the same procedure but with the particle snapshot at $z=0$. 
Finding three eigenvalues, $\{\varrho_{0,1},\varrho_{0,2},\varrho_{0,3}\}$, of ${\bf T}_{0}({\bf x}_{c})$, at the position of each galaxy center, ${\bf x}_{c}$, at $z=0$, 
we compute the local density contrast, $\delta\equiv \varrho_{0,1} + \varrho_{0,2}+\varrho_{0,3}$, and the environmental shear, 
$q\equiv \sqrt{\left[(\varrho_{0,1}-\varrho_{0,2})^{2} + (\varrho_{0,2}-\varrho_{0,3})^{2}+(\varrho_{0,3}-\varrho_{0,1})^{2}\right]/2}$, as well~\citep{par-etal18}. 

For each galaxy in a given main sample, we identify its constituent DM particles from the TNG300-1 particle snapshot at $z=0$. 
Tracing the constituent DM particles of each galaxy at $z=0$ back to $z_{\rm ini}=127$, we locate their initial  comoving positions and determine their center 
of mass, $\bar{\bf x}_{\rm ini}\equiv (x_{{\rm ini},\alpha})$, as a protogalaxy center. 
%%%%%%%%%%%%%%%%%%%%%%%%%%%%%%%%%%%%%%%%%%%%%%%%%%%%%%%%%%%%%%%%
\begin{table}[tbp]
\centering
\begin{tabular}{cccccc}
\hline
\rule{0pt}{4ex} 
$m_{t}$ & $N_{g}$ & $N_{g}(\tau\vert m_{\rm t},\delta)$ & $N_{g}(\tau\vert m_{\rm t},\delta,q)$ & $k$ & $\theta$ \medskip\\
\hline
\rule{0pt}{4ex}    
$[10.5,\ 11.0]$ & 387262 & 134718 & 135312  & 3.33$\,\pm\,$0.01 & 0.013$\,\pm\,$0.0000 \medskip \\
$[11.0,\ 11.5]$ & 148145  & 21006 & 26988  & 3.48$\,\pm\,$0.01 & 0.013$\,\pm\,$0.0001 \medskip \\
$[11.5, \ 13.0]$ & 83491 & 10728 & 13518 & 3.33$\,\pm\,$0.01 & 0.014$\,\pm\,$0.0000 \medskip \\
\hline
\end{tabular}
\caption{\label{tab:ng} Logarithmic mass range, galaxy numbers of the original subsamples, galaxy numbers of the $\tau$-selected and $\delta$-controlled subsamples, 
galaxy numbers of the $\tau$-selected and $\delta$-$q$ controlled subsamples, and best-fit values of $k$ and $\theta$ of the $\Gamma$-distribution that describes $p(\tau)$.}
\end{table}
%%%%%%%%%%%%%%%%%%%%%%%%%%%%%%%%%%%%%%%%%%%%%%%%%%%%%%%%%%%%%%%%
The inertia tensor, ${\bf I}^{\prime}_{\rm ini}\equiv (I_{\rm ini, \alpha\beta})$, of a protogalaxy at the center position of $\bar{\bf x}_{\rm ini}$ is 
determined first in the Cartesian frame~\citep{bet-etal07} from information on the comoving positions of the constituent particles as well as their masses at $z_{\rm ini}=127$: 
%%%%%%%%%%%%%%%%%%%%%%%%%%%%%%%%%%%%%%%%%%%%%%%%%%%%%%%%%%%%%%%%%%%%%%%%%%%%%%%
\begin{equation}
I^{\prime}_{{\rm ini},\alpha\beta}(\bar{\bf x}_{\rm ini})= \sum_{\gamma=1}^{n_{p}}m_{\gamma}\left(x_{{\rm ini},\alpha}[\gamma]-\bar{x}_{{\rm ini},\alpha}\right)
\left(x_{{\rm ini},\beta}[\gamma]-\bar{x}_{{\rm ini},\beta}\right)\, ,
\end{equation} 
%%%%%%%%%%%%%%%%%%%%%%%%%%%%%%%%%%%%%%%%%%%%%%%%%%%%%%%%%%%%%%%%%%%%%%%%%%%%%%%
where the mass and comoving position of the $\gamma$-th constituent particle are denoted by $m_{\gamma}$ and 
${\bf x}_{\rm ini}[\gamma]\equiv (x_{{\rm init},\alpha}[\gamma])$, respectively. The initial tidal tensor at $\bar{\bf x}_{\rm ini}$ is also located and its orthonormal 
eigenvectors, $\{\hat{\bf t}_{{\rm ini},\alpha}\}_{\alpha=1}^{3}$, corresponding to three eigenvalues in a decreasing order are determined for each protogalaxy. 

As done in~\cite{ML24}, we perform a similarity transformation of the inertia tensor to find its expression in the principal frame of ${\bf T}_{\rm ini}(\bar{\bf x}_{\rm ini}; R_{f})$ as 
${\bf I}_{\rm ini} = {\bf W}^{t}\cdot {\bf I}^{\prime}_{\rm ini}\cdot{\bf W}$,  where ${\bf W}$ is a $3\times 3$ rotation matrix with $(W_{\alpha\gamma})=(\hat{\bf t}_{{\rm ini},\alpha})$, 
for $\gamma\in \{1,2,3\}$ and ${\bf W}^{t}$ is the inverse matrix of ${\bf W}$. The inertia tensor, ${\bf I}_{\rm ini}(\bar{\bf x}_{\rm ini})$, 
in the principal frame of ${\bf T}_{\rm ini}(\bar{\bf x}_{\rm ini}; R_{f})$ would have non-zero off-diagonal components, only provided that its three principal axes are not perfectly 
aligned nor anti-aligned with those of ${\bf T}_{\rm ini}(\bar{\bf x}_{\rm ini};R_{f})$. 
For each protogalaxy at $z_{\rm ini}=127$, the degree of misalignment between ${\bf I}_{\rm ini}(\bar{\bf x}_{\rm ini})$ and ${\bf T}_{\rm ini}(\bar{\bf x}_{\rm ini};R_{f})$ is 
determined as 
%%%%%%%%%%%%%%%%%%%%%%%%%%%%%%%%%%%%%%%%%%%%%%%%%%%%%%%%%%%%%%%%%%%%%%%%%%%
\begin{equation}
\tau \equiv \left(\frac{I^{2}_{{\rm ini}, 12}+I^{2}_{{\rm ini}, 23}+I^{2}_{{\rm ini}, 31}}{I^{2}_{{\rm ini},11}+I^{2}_{{\rm ini},22}+I^{2}_{{\rm ini},33}}\right)^{1/2}\, ,
\end{equation}
%%%%%%%%%%%%%%%%%%%%%%%%%%%%%%%%%%%%%%%%%%%%%%%%%%%%%%%%%%%%%%%%%%%%%%%%%%%
where $\{I_{{\rm ini},11},I_{{\rm ini},22},I_{{\rm ini},33}\}$ and $\{I_{{\rm ini},12},I_{{\rm ini},23},I_{{\rm ini},31}\}$ are three diagonal and off-diagonal components 
of ${\bf I}_{\rm ini}(\bar{\bf x}_{\rm ini})$, respectively, in the principal frame of ${\bf T}_{\rm ini}(\bar{\bf x}_{\rm ini})$.  

According to the linear tidal torque theory~\cite{whi84,lp00},  the precondition for the first order generation of a protogalaxy angular momentum ${\bf J}_{\rm ini}$ is nothing 
but a non-zero value of $\tau$.  As mentioned in section~\ref{sec:intro},  \cite{ML24} discovered that the spin orientations of the present galaxies with respect to the principal 
axes of ${\bf T}_{0}(R_{f})$ exhibited sensitive variations with the non-zero values of $\tau$ whose probability density function turned out to be well approximated by the $\Gamma$ 
distribution regardless of $R_{f}$~\cite[see figure 1 in][]{ML24}.  In light of this previous finding, we are going to investigate if $\tau$ affects not only the spin orientations but also the following 
basic traits of the present galaxies: the spin parameters ($\lambda$), formation epochs ($a_{f}$), stellar-to-total mass ratios ($M_{\rm \star}/M_{\rm t}$), and ratios of the 
random motion to total kinetic energy ($K_{\rm rd}/K_{\rm t}$). Here, $a_{f}$ represents the scale 
factor when the main progenitor of a given galaxy garners $M_{\rm t}/2$~\citep{nfw97}, $K_{\rm rd}/K_{\rm t} \equiv 1 - \kappa_\mathrm{rot}$ is the 
fraction of kinetic energyin ordered rotation with $\kappa_\mathrm{rot}$ defined similar to~\cite{sal-etal12} but with all DM and baryonic particles, and $\lambda$ is defined as 
in~\cite{bul-etal01}: $\lambda = J/(\sqrt{2}M_{\rm t}v\,r_{\rm vir})$ where $J$ is the magnitude of the angular momentum measured at the galaxy virial distance $r_{\rm vir}$ from 
the center and $v^{2}\equiv GM_{\rm t}/r_{\rm vir}$.  This choice of the basic galaxy traits is made under the same assumption of \cite{dis-etal08} 
that a galaxy, as a gravitationally bound system, can be well characterized by these four basic traits in addition to its total mass ($M_{\rm t}$).

%%%%%%%%%%%%%%%%%%%%%%%%%%%%%%%%%%%%%%%%%%%%%%%%%%%%%%%%%%%%%%%%
\begin{figure}[tbp]
\centering % \begin{center}/\end{center} takes some additional vertical space
\includegraphics[width=0.85\textwidth=0 380 0 200]{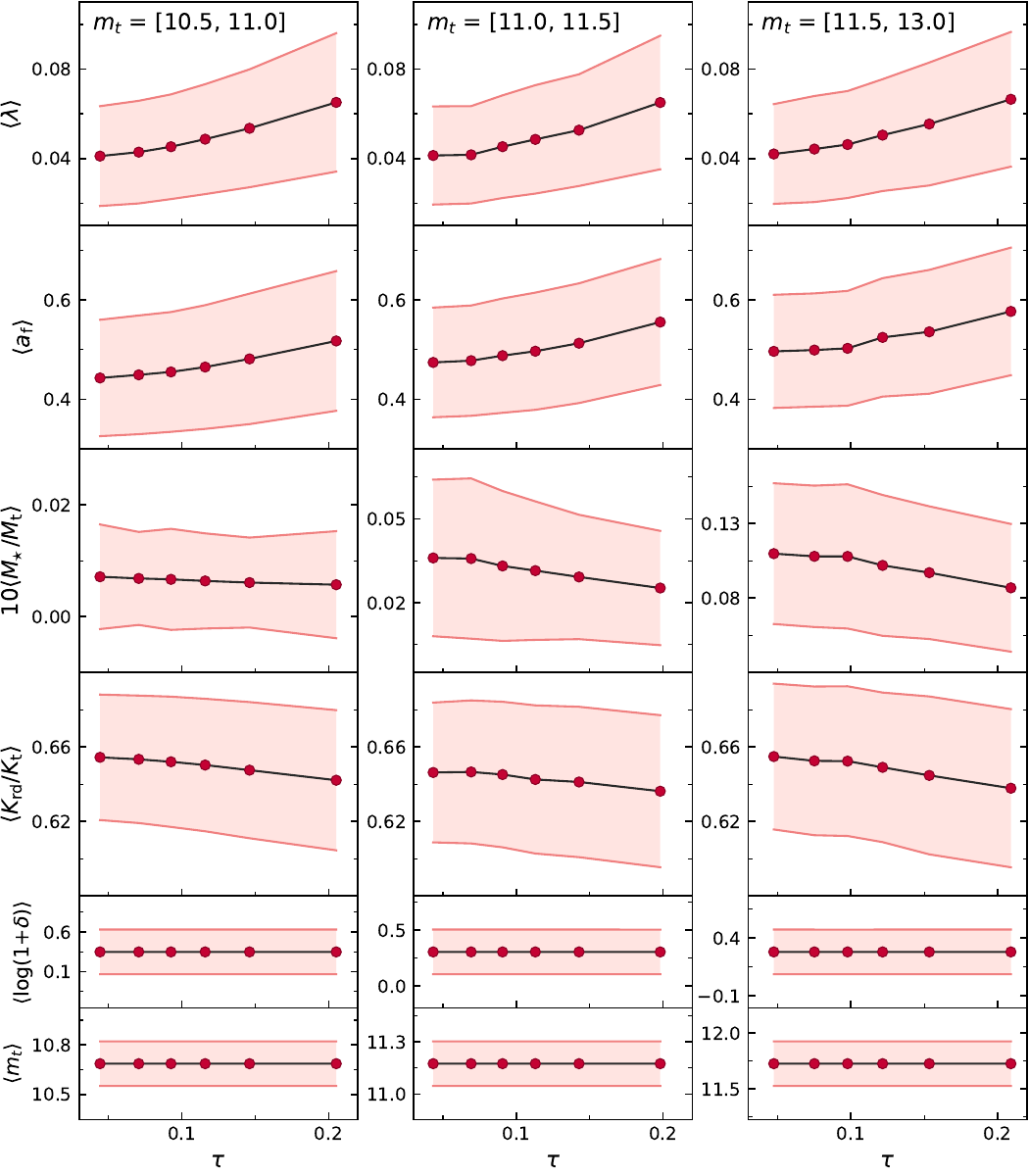}
% "\includegraphics" is very powerful; the graphicx package is already loaded
\caption{\label{fig:ave1} Ensemble averages of the key galaxy traits, logarithmic masses (second panels from the bottom) of the TNG300-1 galaxies, 
and environmental density contrasts (bottom panels), taken over the $\tau$-selected, $m_{t}$ and $\delta$-controlled subsamples in three different $m_{t}$-ranges. 
In each panel, the shaded area corresponds to one standard deviation scatter around the mean values (filled red circles), while the errors in the mean values are too 
small to be visible.}
\end{figure}
%%%%%%%%%%%%%%%%%%%%%%%%%%%%%%%%%%%%%%%%%%%%%%%%%%%%%%%%%%%%%%%%
%%%%%%%%%%%%%%%%%%%%%%%%%%%%%%%%%%%%%%%%%%%%%%%%%%%%%%%%%%%%%%%%
\begin{figure}[tbp]
\centering % \begin{center}/\end{center} takes some additional vertical space
\includegraphics[width=0.85\textwidth=0 380 0 200]{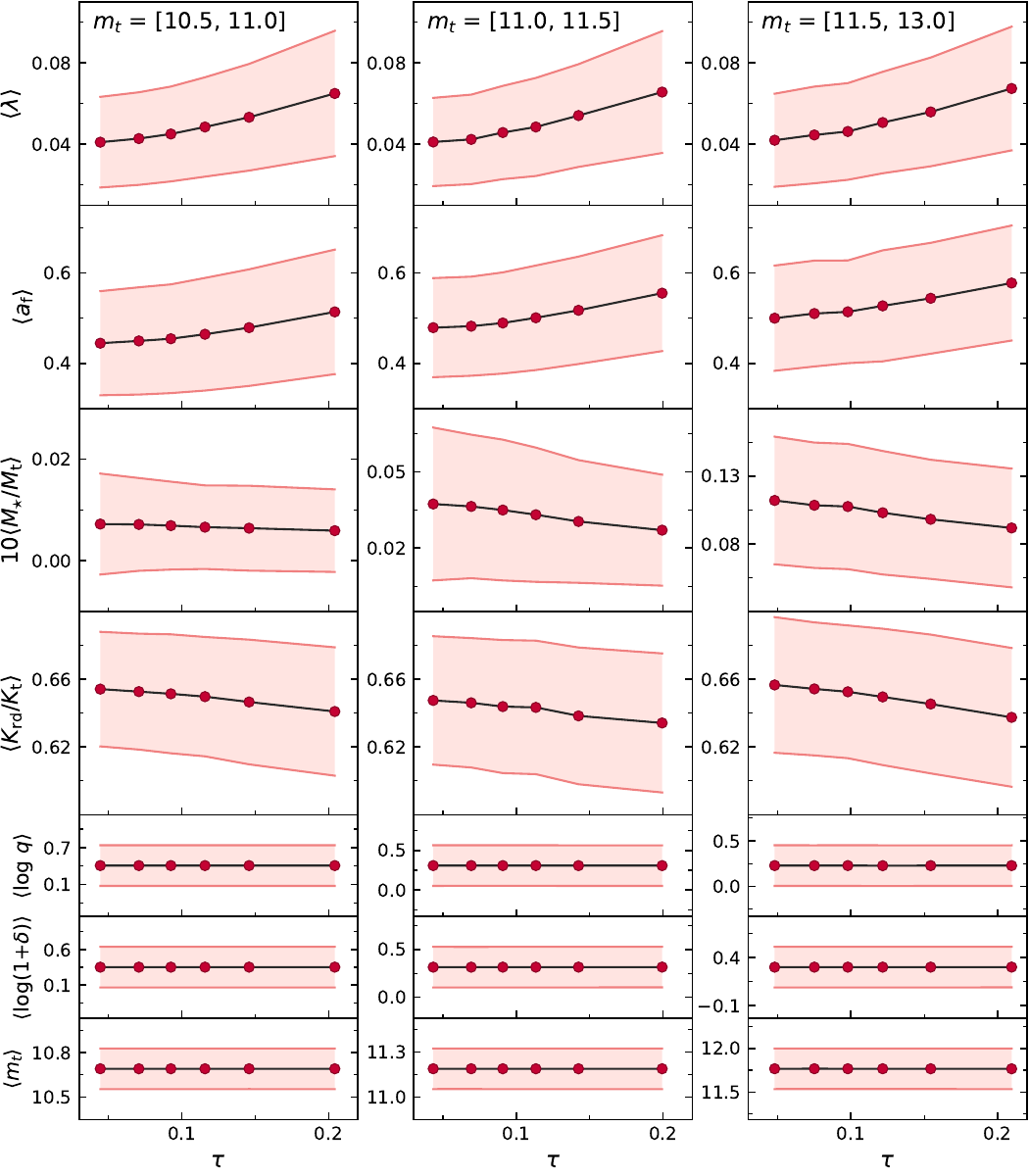}
% "\includegraphics" is very powerful; the graphicx package is already loaded
\caption{\label{fig:ave2} Same as figure~\ref{fig:ave1} but from the $\tau$-selected, $m_{t}$, $\delta$ and $q$-controlled subsamples.}
\end{figure}
%%%%%%%%%%%%%%%%%%%%%%%%%%%%%%%%%%%%%%%%%%%%%%%%%%%%%%%%%%%%%%%%

\subsection{Trends of the $\tau$-dependences of the average basic traits}\label{sec:trend}

We split the first main sample of the galaxies with $m_{t}\ge 11.5$ into six $\tau$-selected subsamples of comparable sizes, and then control them to have no difference in the joint 
distributions of $m_{t}$ and $\log(1+\delta)$ among them via the following procedure:   
(i) The two-dimensional configuration space spanned by $m_{t}$ and $\log(1+\delta)$ is divided into multiple pixels. (ii) The numbers of the galaxies from each of the six $\tau$-selected 
subsamples are counted and its minimum value ($n_{\rm min}$) is found at a given pixel. (iii) Only $n_{\rm min}$ galaxies are extracted from each of the six $\tau$-selected to constitute 
a new controlled subsample. This procedure is required to single out the effect of $\tau$ from those of $m_{t}$ and $\delta$ on the galaxy traits
\citep[e.g.,][]{PG84,gom-etal03, tan-etal04,ree-etal05,wan-etal11,lee-etal17,chi-etal21}. 

The ensemble averages of the aforementioned basic traits of the galaxies and one standard deviation scatters  are taken over each of the $\tau$-selected, $m_{t}$ and 
$\delta$-controlled subsamples as $\langle \eta\rangle = \sum_{i=1}^{N_{g}} \eta\,/\,N_{g}$ and $\sigma^{2}_{\eta}=\sum_{i=1}^{N_{g}}\left(\eta-\langle\eta\rangle\right)^{2}\,/\,N_{g}$ 
with $\eta\in\{\lambda,a_{f},M_{\star}/M_{\rm t}, K_{\rm rd}/K_{\rm t}\}$ with number of the galaxies $N_{g}$ at each $\tau$-bin. 
The right columns of figure~\ref{fig:ave1} plot $\langle \eta\rangle$ and $\sigma_{\eta}$ as filled red 
circles and shaded area, respectively. The errors in the ensemble averages, $\sigma_{\eta}/\sqrt{N_{g}-1}$, are also plotted but too small to be visible due to the high values 
of $N_{g}$.  The bottom two rows plot the ensemble averages of $m_{t}$ and $\log(1+\delta)$, revealing that the six $\tau$-selected, $m_{\rm t}$ and $\delta$-controlled 
subsamples indeed have no differences in the total masses and environmental densities. 

As can be seen,  the ensemble average of each galaxy trait shows a clear trend of $\tau$-dependent variation. 
The increment of $\tau$ leads $\langle\lambda\rangle$ and $\langle a_{f}\rangle$ to increase, while $\langle M_{\star}/M_{\rm t}\rangle$ and $\langle K_{\rm rd}/K_{\rm t}\rangle$ 
to decrease.  We repeat the same analysis but with the other two main samples of the galaxies in the two lower mass ranges, the results of which are shown in the 
left and middle columns of figure~\ref{fig:ave1}. The number of the galaxies belonging to the $\tau$-selected, $m_{\rm t}$ and $\delta$-controlled subsamples in the three mass
ranges are listed in the third column of table~\ref{tab:ng}. Note that since the six subsamples are all controlled, they share the same numbers of the galaxies in each $m_{t}$-range.

 As can be seen, the galaxies in these lower mass ranges exhibit the same trends of $\tau$-dependent variations of 
their traits.  As can be seen, the three traits, $\langle\lambda\rangle$, $\langle a_{f}\rangle$, and $\langle K_{\rm rd}/K_{\rm t}\rangle$ exhibit similar $\tau$-dependent 
variations even in the lower mass ranges.  Meanwhile, the other stellar trait, $\langle M_{\star}/M_{\rm t}\rangle$, shows very little $\tau$-dependent variation in the lowest-mass 
range of $10.5\le m_{t}< 11$. This result implies that the stellar traits of low-mass galaxies should be more vulnerable to the nurture effect of subsequent evolutions, which plays 
the role of wiping out their initial memory for $\tau$. 

Given that the local environmental shear is also known to have an independent effect on the galaxy properties~\citep{par-etal18,HP88,CT96,LL08}, 
the $\tau$-subsamples are further controlled to have an identical joint distribution of $m_{t}$, $\delta$ and $q$. The ensembles averages of the galaxy basic traits 
taken over the $\tau$-selected and $m_{t}$, $\delta$ and $q$-controlled subsamples are shown in figure~\ref{fig:ave2}, and the numbers of the galaxies belonging to each subsample 
are listed in the fourth column of table~\ref{tab:ng}. As can be seen, the four basic traits of the galaxies yield very similar trends, despite that no differences in $q$ among the six 
subsamples exist. 
The results shown in figures~\ref{fig:ave1}-\ref{fig:ave2} imply that the galaxies originated from the protogalactic sites whose inertia tensors are more strongly misaligned with 
the initial tidal tensors have an average tendency of spinning faster, forming later, less luminous, containing higher fraction of kinetic energy
in ordered rotation, and thus more rotationally supported, regardless of their masses, environmental densities and shears. 

\subsection{Mutual information between $\tau$ and the basic galaxy traits}\label{sec:MI}

Although the ensemble averages of the four basic traits from the $\tau$-selected, mass and environmental effects controlled subsamples allow us to see the overall trends by 
which they vary with $\tau$,  the large scatters around their averages shown in figure.~\ref{fig:ave1}-\ref{fig:ave2} deter us from making a conclusive assessment of 
the causal impact of $\tau$ on the key galaxy traits. For a causal inference, it may be more useful to quantify the $\tau$-dependence of the galaxy traits in terms of the mutual 
information (MI), defined as \citep{sha48,SP20}
%%%%%%%%%%%%%%%%%%%%%%%%%%%%%%%%%%%%%%%%%%%%%%%%%%%%%%%%%%%%%%%%%%%%%%%%%%%%%%%%%%%%%%%%%%%%
\begin{equation}
\label{eqn:mutual}
{\rm MI}(\tau, \eta)= \int\,d\tau\int\,d\eta\,p(\tau,\eta)\log\left[\frac{p(\tau,\eta)}{p(\tau)p(\eta)}\right]\, ,
\end{equation}
%%%%%%%%%%%%%%%%%%%%%%%%%%%%%%%%%%%%%%%%%%%%%%%%%%%%%%%%%%%%%%%%%%%%%%%%%%%%%%%%%%%%%%%%%%%%
where both of $\tau$ and $\eta$ are regarded as random variables, $p(\tau)$ and $p(\eta)$ are the one-point probability functions of $\tau$ and $\eta$, respectively, 
while $p(\tau,\eta)$ is the joint probability function of a random vector $(\tau,\eta)$.  Since the MI is a measure of the {\it similarity} (or equivalently mutual dependence) 
between the two random variables in terms not of their averages but of the entire shapes of their probability distributions \citep{sha48,SP20}, 
it is more rigorous and quite advantageous for the quantification of the impact of one random variable on the other, especially when their correlations suffer from large scatters. 
If $\tau$ and $\eta$ have no relationship with each other, then the ${\rm MI}$ will be zero. The more strongly $\eta$ is dependent on $\tau$, the more significantly the MI will  
deviate from zero. 
 
To compute the MI between $\tau$ and a galaxy trait $\eta$ with $\eta\in\{\lambda, a_{f}, M_{\star}/M_{\rm t}, K_{\rm rd}/K_{\rm t}\}$, 
we first determine three probability functions: Splitting the entire $\tau$-range into multiple short intervals of equal length $\Delta_{\tau}$, 
and counting the number, $n_{g}(\tau)$, of the galaxies at a given $\tau$-bin, we determine $p(\tau)$ as $p(\tau)=n_{g}(\tau)/(\Delta_{\tau}N_{g})$. 
As mentioned in section~\ref{sec:tau}, it was found by~\cite{ML24} that $p(\tau)$ is well approximated by the $\Gamma$-distribution. 
In a similar manner, we also obtain $p(\eta)$.  
For the evaluation of $p(\tau,\eta)$,  the two-dimensional area spanned by $\tau$ and $\eta$ is divided into multiple pixels of equal area $\Delta_{\tau\eta}$ 
and the number of the galaxies, $n_{g}(\tau,\eta)$, at each pixel is counted to determine $p(\tau, \eta)$ as $p(\tau,\eta)\equiv n_{g}(\tau,\eta)/(\Delta_{\tau\eta}N_{g})$. 

\begin{figure}[tbp]
\centering % \begin{center}/\end{center} takes some additional vertical space
\includegraphics[width=0.85\textwidth=0 380 0 200]{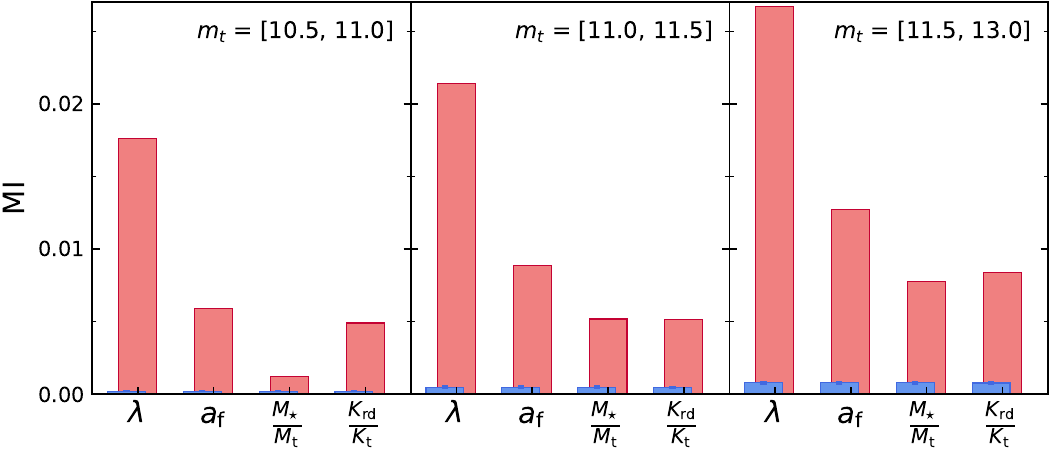}
% "\includegraphics" is very powerful; the graphicx package is already loaded
\caption{\label{fig:mi} Mutual informations about $\tau$ (filled red histograms) shared by the basic traits of the galaxies  
in the three mass ranges. Mean MI's averaged over $1000$ resamples created by shuffling galaxies (filled blue histograms) 
and $1\sigma$ scatters (blue error bars) are shown for comparison.}
\end{figure}

Plugging these probability functions into eq.~(\ref{eqn:mutual}), we compute the {\rm MI} between $\tau$ and each of the basic traits of the galaxies 
belonging to the three main samples, the results of which are shown as filled red histograms in figure~\ref{fig:mi}. 
Randomly and repeatedly shuffling the positions of the galaxies of each main sample, we create $1000$ resamples over each of which the amount of MI between $\tau$ and $\eta$ 
is recalculated.  The average and one standard deviation scatter of the MI are taken over the $1000$ resamples of the shuffled galaxies, which are shown as 
blue histograms with error bars in figure~\ref{fig:mi}.  In each panel the error bars represent not $1\sigma$ in the mean MI but $1\sigma$ scatter around the mean among 
the $1000$ resamples. 

As can be seen, each of the basic galaxy traits and initial precondition $\tau$ indeed share a significantly large amount of MI with each other. 
The largest amount of MI about $\tau$ is found to be possessed by $\lambda$, which is in fact naturally expected given that the spin parameter of a present 
galaxy should be most closely linked with the initial precondition of protogalaxy angular momentum. 
Meanwhile, the smallest MI with $\tau$ is exhibited by the stellar property, $M_{\star}/M_{\rm t}$, which may be related to the stochastic nature of galaxy star formation. 
As shown by \cite{CT19,tac-etal20,wan-etal22,iye-etal24}, the star formation histories are stochastically different among individual galaxies,  which are caused not only 
by unknown physical factors in the stellar evolutions of galaxies but also by the numerical butterfly effect~\cite[see][for details]{gen-etal19}. 

Note that the more massive galaxies tend to contain higher and more significant amounts of MI about $\tau$. 
This result is consistent with the numerical findings~\cite{ML23,ML24} that the spin directions of more massive galaxies tend to retain better the initially acquired alignment tendencies 
with the principal axes of the linear tidal field.  As shown in~\cite{ML23}, the orbital angular momentum transfer during gravitational merging events leads the spin angular 
momentum of a merged galaxy to develop a strong correlation with a larger scale (less nonlinear) tidal field. The more frequent mergers a massive galaxy experiences, 
the better memory its final spin possesses for the initial tidal field in the linear regime.  
Furthermore, for the case of relatively poorly resolved low-mass galaxies, the measurements of their inertia tensors suffer from larger Poisson noise~\cite{bet-etal07}, 
which must also contribute to their low amounts of MI about $\tau$. 

It is worth mentioning here that the significance of a MI signal is estimated by the degree of its deviation from the value obtained from randomly shuffled samples. 
Unlike the simple linear correlation parameter, the actual value of MI between two random variables could be quite small even if they are strongly 
correlated with each other, since MI is determined by the full shapes of their probability density functions.  In other words, 
unless their ranges and probability density functions are identical, the face value of MI between two random variables could be low, no matter how strong 
the mutual interdependence is. 

\begin{figure}[tbp]
\centering % \begin{center}/\end{center} takes some additional vertical space
\includegraphics[width=0.85\textwidth=0 380 0 200]{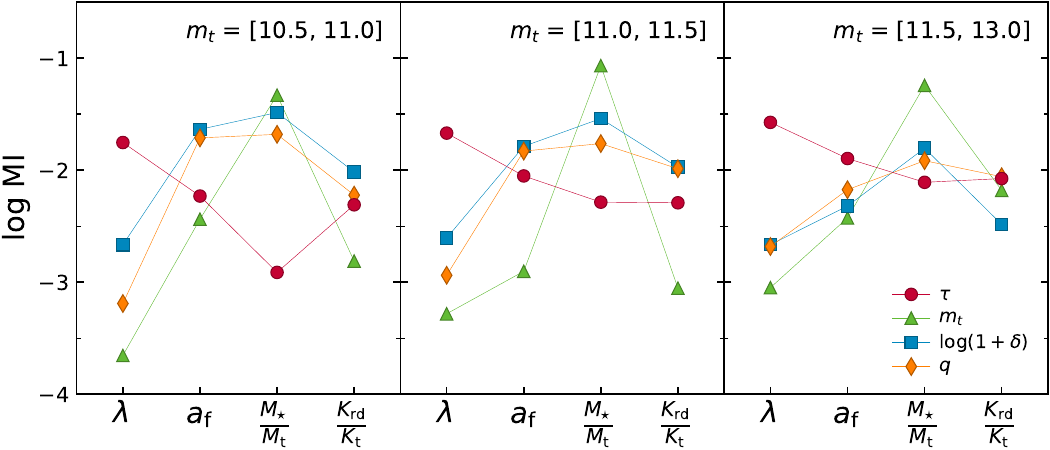}
% "\includegraphics" is very powerful; the graphicx package is already loaded
\caption{\label{fig:mi_com} Comparisons among the mutual informations shared by the galaxy basic traits in the three mass ranges 
about $\tau$ (filled red circles), $M_{\rm t}$ (filled blue squares), $\delta$ (filled green triangles) and $q$ (filled orange diamonds).}
\end{figure}

We also compute ${\rm MI}(M_{\rm t},\eta)$, ${\rm MI}(\delta,\eta)$ and ${\rm MI}(q,\eta)$ by replacing $\tau$ by $m_{t}$, $\log (1+\delta)$ and $q$, respectively, in eq.~\ref{eqn:mutual}, 
the results of which are shown (filled green triangles, blue squares and orange diamonds, respectively) and compared with ${\rm MI}(\tau, \eta)$ (filled red circles) 
in figure~\ref{fig:mi_com}. As can be seen, for the case of $\eta=\lambda$, ${\rm MI}(\tau,\lambda)$ is the largest over 
${\rm MI}(M_{\rm t},\lambda)$, ${\rm MI}(\delta,\lambda)$ and ${\rm MI}(q,\lambda)$ regardless of the $m_{t}$-ranges, 
which indicates that $\tau$ has the strongest impact on the galaxy spin parameters than the total mass and environments.
For the case of $\eta=M_{\star}/M_{\rm t}$, however,  ${\rm MI}(\tau,M_{\star}/M_{\rm t})$ is the smallest in all of the three $m_{t}$ ranges. 
The $\tau$-dependence of the stellar to total mass ratios of the galaxies, albeit significant, is not comparable in strength to its dependences on the total mass and environmental 
shear. 

For the cases of $\eta=a_{f}$,  ${\rm MI}(\tau,a_{f})$ is the largest in the mass range of $11.5\le m_{t}< 13$, indicating that the initial precondition $\tau$ plays a more decisive role 
than the galaxy masses and environmental shear in determining when the giant galaxies form.
This result is somewhat surprising, challenging the conventional assumption based on the hierarchical merging scenario that the formation epochs of the galaxies are nurtured by 
the environmental factors at fixed masses since the hierarchical merging rates depend strongly on the environmental densities~\cite{LC93,LC94}. 
Meanwhile, in the lower mass ranges of $10.5\le m_{t}< 11.5$, ${\rm MI}(\tau,a_{f})$ is larger than ${\rm MI}(M_{\rm t},a_{f})$ but smaller than ${\rm MI}(\delta,a_{f})$ and 
${\rm MI}(q,a_{f})$. Although $\tau$ does not play the most dominant role in determining when the lower mass galaxies form, its effect seems to be stronger than the total masses.  
For the cases of $\eta=K_{\rm rd}/K_{\rm t}$, a similar $M_{\rm t}$-dependent dominance is found: ${\rm MI}(\tau,K_{\rm rd}/K_{\rm t})$ is the largest in the highest mass range, 
while ${\rm MI}(M_{\rm t},K_{\rm rd}/K_{\rm t})<{\rm MI}(\tau,K_{\rm rd}/K_{\rm t})< {\rm MI}(q,K_{\rm rd}/K_{\rm t})\le {\rm MI}(\delta,K_{\rm rd}/K_{\rm t})$ 
in the lower mass ranges. 

%%%%%%%%%%%%%%%%%%%%%%%%%%%%%%%%%%%%%%%%%%%%%%%%%%%%%%%%%%%%%%%%
%%%%%%%%%%%%%%%%%%%%%%%%%%%%%%%%%%%%%%%%%%%%%%%%%%%%%%%%%%%%%%%%
%%%%%%%%%%%%%%%%%%%%%%%%%%%%%%%%%%%%%%%%%%%%%%%%%%%%%%%%%%%%%%%%
%%%%%%%%%%%%%%%%%%%%%%%%%%%%%%%%%%%%%%%%%%%%%%%%%%%%%%%%%%%%%%%%

\section{Summary and conclusion}\label{sec:con}

An application of Shannon's information theory~\cite{sha48} to the IllustrisTNG300-1 simulation data ~\citep{tng1,tng2,tng3,tng4,tng5,tng6} has revealed that the initial precondition for 
protogalaxy angular momentum, $\tau$, quantified by the degree of misalignments between the protogalaxy inertia and initial tidal tensors at $z_{\rm ini}=127$, shares a significantly large 
amount of mutual information (MI) with each of the following basic traits of the galaxies in a wide range of logarithmic mass, $10.5\le m_{t}\equiv\log M_{\rm t}/(h^{-1}M_{\odot})\le 13$ at $z=0$: 
the spin parameter ($\lambda$), scale factor at the formation epoch ($a_{f}$), stellar to total mass ratio ($M_{\star}/M_{\rm t}$), and fraction of kinetic energy
in ordered rotation ($K_{\rm rd}/K_{\rm t}$).  
Interpreting the amount of MI as the strength of impact, we have obtained somewhat surprising, unexpected result that the initial condition $\tau$ is even more impactful 
than the total masses and environments on some of the basic traits: For the case of the high-mass galaxies with $m_{t}\ge 11.5$,  $\tau$ is more impactful 
than the total masses and environments on the three basic traits,  $\lambda$, $a_{f}$, and $K_{\rm rd}/K_{\rm t}$, while for the case of the dwarf and low-mass galaxies with $m_{t}< 11.5$, 
it is true only on $\lambda$.  It has also been demonstrated that in the entire mass range considered, $\tau$ is more impactful than the total masses on $\lambda$ and $a_{f}$. 
Whereas, the impact of $\tau$ on $M_{\star}/M_{\rm t}$ has turned out to be weaker than those of the total masses and environments.  
It has also been found that the probability density function of $\lambda$ is well approximated by the $\Gamma$-distribution, just like $\tau$~\cite{ML24}. 

To see the direction of a trend by which each of the four traits changes with $\tau$, we have computed its ensemble average as a function of $\tau$ by separating the galaxies into 
multiple $\tau$-selected subsamples all of which are controlled to have no difference in the joint distributions of total mass, environmental density and shear.  
It has been shown that $\lambda$ and $a_{f}$ tend to increase with $\tau$, while the other two traits exhibit an opposite trend. 
The consequential implication of our findings is as follows:  A galaxy condensed out from the protogalaxy site with a higher $\tau$ value tends to develop a faster spinning motion 
characterized by a higher spin parameter, which in turn has an effect of retarding its formation (increasing $a_{f}$),  hindering the infall and cooling of gas particles leading to a low 
stellar to mass ratio~\citep{lu-etal22}, and decreasing the fraction of its random motion kinetic energy~\citep{jim-etal98,KL13,cad-etal22}. 
These results support our hypothesis that the basic traits of the present galaxies are still connected with the initial states, which is likely to contribute to lowering the dimension of 
their correlation structure as observed by~\cite{dis-etal08}. 

Our results might also provide crucial clues to several unexpected features of the observed or simulated galaxies. For instance, let us recall the question recently raised 
by~\cite{nai-etal10}: 
why the luminosity-size scaling relation of the Sloan Digital Sky Survey (SDSS) galaxies shows no correlation with environmental density although the luminosity itself exhibits strong 
environmental variation.  Recalling that the galaxy sizes are closely related to the spin parameters, which have been found in the current work to be most significantly dependent on $\tau$, 
we speculate that the observed luminosity-size scaling relation may well reflect the initial precondition $\tau$, independent of the local environments. 
Another example is the result of~\cite{cad-etal22} from a hydrodynamical zoom-in simulation of three galaxies that the stellar angular momenta are regulated largely by the initial 
conditions rather than by any evolutionary effects. In their work, the angular momenta of protogalaxies at $z=200$ were deliberately controlled to show that the initial conditions 
control the angular momenta, sizes and bulge fractions of the stellar components of the present galaxies, even though the star formation histories of individual galaxies are 
highly stochastic~\cite{CT19,tac-etal20,wan-etal22,iye-etal24,gen-etal19}.
Our results naturally answer the question of what the initial condition should be. 

When the low dimensional correlation structure from the galaxy properties was first detected by the observational study of~\cite{dis-etal08}, 
it was thought of as an observational challenge to the currently favored CDM paradigm according to which the galaxies evolve through hierarchical merging processes. 
Not to mention the stochastic nature of star formation histories~\cite{CT19,tac-etal20,wan-etal22,iye-etal24,gen-etal19}, the hierarchical merging histories themselves are expected to be 
haphazardly differentamong the galaxies, nurturing them to have a variety of dissimilar properties. 
This conventional expectation based on the CDM paradigm made it very hard to envision the existence of such a simple correlation structure of the key properties of the galaxies 
with comparable masses. Our work has, however, disproved this conventional expectation, indicating compatibility of the existence of this simple correlation structure with the CDM 
paradigm if the initial precondition $\tau$ is properly taken into account, and implying that the galaxy traits are molded at least partially by 
nature, that is, by cosmological predispositions. 

Now that $\tau$ is unveiled as a hidden player in the act of galaxy formation and evolution, we expect that our results may shed a whole new insight on the halo occupation number 
statistics~\citep{BW02} as well as on the origin of halo assembly~\citep{GW07}. Moreover, our result hints at the possibility that the directly observable properties of the present galaxies 
might be useful as a complimentary near-field probe of the initial conditions of the universe.  For instance, by directly measuring the galaxy spin parameters from the local universe, 
we should be able to extract a substantially large amount of information on the distribution of the initial precondition, $\tau$, which in turn may well depend on the background cosmology. 
We plan to address these issues in the future and hope to report a result elsewhere. 

\appendix
\section{Modeling $p(\lambda)$ by the $\Gamma$-distribution}

\begin{figure}[tbp]
\centering
\includegraphics[width=0.85\textwidth=0 380 0 200]{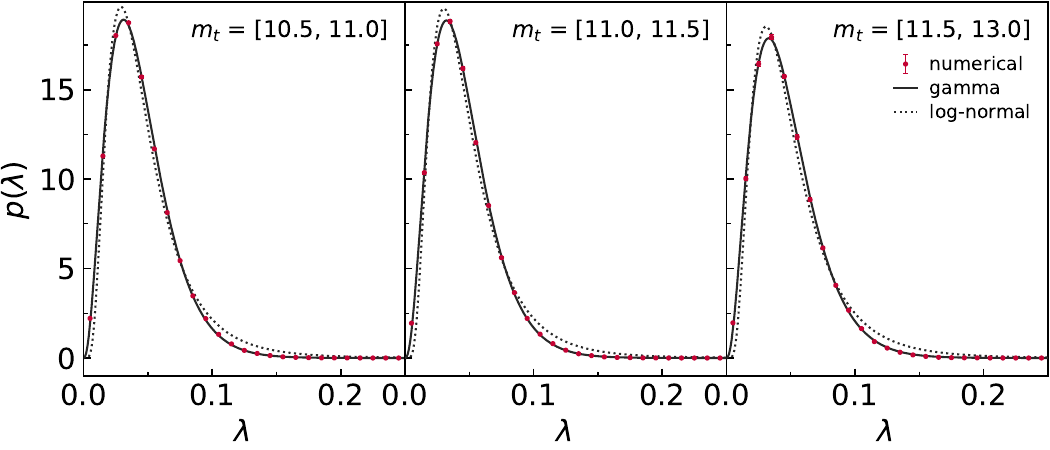}
\caption{\label{fig:lam} Probability density functions of the spin parameters of the TNG300-1 galaxies (red filled circles) 
compared with the best-fit $\Gamma$-distributions (black solid lines) as well as with the best-fit 
log-normal counterparts (black dotted lines).}
\end{figure}

Recalling that the probability density function of $p(\tau)$ was found to be well approximated by the $\Gamma$ distribution~\cite{ML24} and noting that the spin parameter $\lambda$ shares the 
largest amount of mutual information with $\tau$ (section~\ref{sec:MI}), we explore if $p(\lambda)$ may also be better modeled by the following $\Gamma$-distribution than by the 
conventional log-normal distribution \citep{bul-etal01}:
%%%%%%%%%%%%%%%%%%%%%%%%%%%%%%%%%%%%%%%%%%%%%%%%%%%%%%%%%%%%%%%
\begin{equation}
\label{eqn:gam}
p(\lambda;k,\theta) = \frac{1}{\Gamma(k)\theta^{k}}\lambda^{k-1}\exp\left(-\frac{\lambda}{\theta}\right)\, .
\end{equation}
%%%%%%%%%%%%%%%%%%%%%%%%%%%%%%%%%%%%%%%%%%%%%%%%%%%%%%%%%%%%%%%
where $k$ and $\theta$ are two free parameters, while $\Gamma(k)\equiv \int_{0}^{\infty} x^{k-1}e^{-x}\,dx$. Figure~\ref{fig:lam} plots the numerically obtained probability 
density functions of $\lambda$ (filled red circles) in the three mass ranges, and compares them with the best-fit $\Gamma$-distributions (black solid lines) as well as with the best-fit log-normal 
distributions (black dotted lines). The $\chi^{2}$-minimization method is employed to determine the best-fit parameters of both of the distributions. 
As can be seen, the $\Gamma$-distribution given in eq.~(\ref{eqn:gam}) yields a much better agreement with the numerical result of 
$p(\lambda)$ than the log-normal counterpart in each mass range. The best-fit parameters, $k$ and $\theta$, are shown in the fifth and sixth columns of table~\ref{tab:ng}. 
This result implies that $\lambda$ shares with $\tau$ not only the largest MI but also the same shape of probability distribution. 
%%%%%%%%%%%%%%%%%%%%%%%%%%%%%%%%%%%%%%%%%%%%%%%%%%%%%%%%%%%%%%%%
%%%%%%%%%%%%%%%%%%%%%%%%%%%%%%%%%%%%%%%%%%%%%%%%%%%%%%%%%%%%%%%%

\acknowledgments

The IllustrisTNG simulations were undertaken with compute time awarded by the Gauss Centre for Supercomputing (GCS) 
under GCS Large-Scale Projects GCS-ILLU and GCS-DWAR on the GCS share of the supercomputer Hazel Hen at the High 
Performance Computing Center Stuttgart (HLRS), as well as on the machines of the Max Planck Computing and Data Facility 
(MPCDF) in Garching, Germany.  JSM acknowledges the support by the National Research Foundation (NRF) of Korea grant 
funded by the Korean government (MEST) (No. 2019R1A6A1A10073437). JL acknowledges the support by Basic Science 
Research Program through the NRF of Korea funded by the Ministry of Education (No.2019R1A2C1083855). 


\end{document}